# Cyber Human Interaction

Michael A. Rupp[1] and Aman Behal[2]

1: Department of Psychology, University of California, Riverside  2: Department of Electrical and Computer Engineering and the NanoScience Technology Center, University of Central Florida.

**Article Outline**

Glossary

Definition of the Subject

Introduction

Wheelchair Mounted Robotic Arms

Characteristics of Users of Assistive Robotic Technology

Older Adults

Instrumental Activities of Daily Living

User Acceptance of Assistive Robots

Individual Differences in User Acceptance of Assistive Robotics

The Need an Adaptive Indirect HRI Framework for Assistive Robotics

Development of a Human Performance Model with Assistive Robotic Devices

Age Differences in Performance Using a WMRA

Gender Differences in Performance Using a WMRA

Types of Human Robot Interaction, Levels, and Types of Automation

Future Directions

## Glossary

**Human Robot Interaction (HRI).** A field of study devoted to the understanding of the design and evaluation of robotic systems designed for use by humans. Additionally, a broad term for the general use, interaction or communication between robotic systems and humans.

**Instrumental Activities of Daily Living (IADLs).** Every day activities, not necessary for sustaining life, but allow an individual to live independently and maintain their well-being (e.g., household chores, preparing food, grooming, leisure activities, etc.).

**Compensation.** Modifications and Processes, to the robot, task, or operating environment, completed in order to reduce the impact of a mobility, physical, sensory, or cognitive limitation.

**Framework.** A set of structured concepts along with their definitions designed based on relevant scholarly literature used to explain a process or topic.

**Wheelchair Mounted Robotic Arm (WMRA).** A type of portable assistive robotic device designed to be attached and work with a powered wheelchair or scooter, consisting of a base, arm segments, and an end-effector (such as a gripper, hook, or other apparatus designed to interact with the environment).

**Time on Task (ToT).** One common performance metric used to describe performance and calculated from the onset of movement of the robot to the end of the particular task.

**Number of Moves (NoM).** One common performance metric used to describe performance and calculated by the total number of times the robotic operator (user) initiates a movement along the Cartesian plane (x,y,z directions).

**Number of Moves per Minute.** One common performance metric used to describe performance and measured by the average number of movements completed in a minute adjusted for total movement duration.

**Processing Speed.** A fluid intellectual ability referring to how quickly an individual can process sensory information and when necessary, select and carry out a response to those stimuli.

**Robotic Trust.** The belief, similar to trust between people, that a robotic system is supportive and capable of user goals in situations where the user cannot have complete knowledge of the outcomes of the actions or performance of the robotic system.

**Spatial Visualization.** One facet of spatial ability, often called mental rotation, in which changes in the orientation of objects are visualized using mental imagery independent of the viewpoint of the observer.

**Spatial Orientation.** One facet of spatial ability, often called perspective taking, in which the changes in the orientation of the environment are visualized using mental imagery dependent on the viewpoint of the observer.

**Working Memory Capacity.** A fluid intellectual ability referring to the limit of chunks or groups of items which can be mentally processed and acted upon simultaneously.

**Gross Dexterity.** The physical ability to perform large or gross manipulations with the hands and fingers which may include grabbing, turning, twisting or other movements.

**Fine Dexterity.** The physical ability to perform small or fine manipulations with the hands and fingers.

**Depth Perception.** Sensitivity to the differences between the visual images created by the eyes. The ability to utilize depth information presented in the environment is also called stereopsis.

**Visual Acuity.** The ability to discern visual detail at a given distance often presented as a ratio of how much detail a particular observer can discern at a distance of 20 feet compared to the distance at which an observer with normal vision could discern the same amount of detail. For example, a score of 20/100 indicates that this particular observer can discern the same amount of detail standing 20 feet away, as an observer with normal vision standing at a distance of 100 feet. This calculation is used to generate weak and strong vision scores.

**Random Forest Models.** A machine learning computer science technique for classification. This technique constructs multiple decision trees and using a bootstrap methodology, variables are tested against random samples of data derived from the existing data to determine which variables improve the model's predictive ability.

## Definition of the Subject

Cyber human interaction is a broad term encompassing the range of interactions that humans can have with technology. While human interaction with fixed and mobile computers is well understood, the world is on the cusp of ubiquitous and sustained interactions between humans and robots. While robotic systems are intertwined with computing and computing technologies, the word robot here describes technologies that can physically affect and in turn be affected by their environments which includes humans. This chapter delves into issues of cyber human interaction from the perspective of humans interacting with a subset of robots known as assistive robots. Assistive robots are robots designed to assist individuals with mobility or capacity limitations in completing everyday activities, commonly called instrumental activities of daily living (IADLs). IADLs range from household chores, eating or drinking to any activity with which a user may need the daily assistance of a caregiver to complete. One common type of assistive robot is the wheelchair mounted robotic arm (WMRA). This device is designed to attach to a user's wheelchair to allow him or her to complete their IADLs independently. In short, these devices have sensors that allow them to sense and process their environment with varying levels of autonomy to perform actions that benefit and improve the well-being of people with capability limitations or disabilities. While human-robot interaction is a popular research topic, not much research has been dedicated with regard to individual with limitations. Due to this, several questions regarding these devices remain unanswered. For example, how much autonomy should a robot have? What are the optimal ways for a robot to interact with a user? How should information be presented to the user by the robot? In this chapter, we provide an overview of assistive robotic devices with emphasis on the WRMA, discuss common methods of user interaction, and the need for an adaptive compensation framework to support potential users in regaining their functional capabilities through use of WMRA devices.

## Introduction

According to the International Federation of Robotics (IFR), there were between 1.2 and 1.5 million industrial robots in operation worldwide at the end of 2011 [1]. While it is difficult to estimate the total number of service robots in worldwide operation, about 3 million personal and domestic service robots and 16,067 professional service robots were sold in 2012 alone. While 94,800 additional professional service robot units will be sold during 2013-16, it is expected that an additional 22 million personal service robots will be sold during the same time period. The market for assistive robots is currently small, but growing. The report notes that numerous national research projects in many nations are focused on this huge future market for assistive high-tech service robots [1].

Humans have a long history of utilizing robots for assistance in completing everyday activities, perhaps inspired by science fiction examples such as Rosie from the Jetsons. However, the earliest realistic assistive robotic systems looked nothing like their fictional counterparts. Instead they were large workstations where users could visit and be provided assistance in completing eating, personal hygiene, and completing limited office work [2]. Later these devices were miniaturized into tabletop [3], mobile devices [4] and those that can attach to a wheelchair [5] allowing for increased design flexibility and increased functional capability for the user. Today, a variety of assistive robotic devices/arms have been developed to improve functional ability assist users in the performance of IADLs [6]. Some or all of these tasks may be completed by a relative, nurse, or various other caregivers [7]; however, technology, namely the field of assistive robotics, has the ability to allow individuals to regain some autonomy and competence by completing activities that would otherwise require assistance from others to complete. Previously, technological devices such as walkers or prosthetics have been used to increase mobility, and devices such as glasses and hearing aids have been used to augment the senses [7]. Many of these assistive devices are static and only perform a limited number of functions. On the other hand, assistive robotic devices are a form of adaptable and intelligent technology that can assist users to complete a wide variety of IADLs based on individuals' specific needs [8]. Additionally, due to their versatility, these devices have great potential to improve individuals' quality of life through increasing their functional independence [9].

Several assistive robots have been created for use in research and commercially. Specifically, the Desktop Robot has been demonstrated for upper-arm tasks [10]-[15]. Users have also responded well to specialized assistive devices known as Powered Aids

[14]–[16], e.g., MySpoon and Handy (feeding aids), Dorma (door opener), Turny (page turner), and Harrison (curtain opener). However, these devices are large, heavy and non-mobile, thus a user's home may be equipped with many of these devices each to accomplish a single task. To overcome this limitation inherent with Powered Aids and Desktop Robots, more vagile robots were needed. Specifically, several wheelchair-mounted robot arms (WMRAs) have been proposed. KARES I [17], FRIEND [18], ARM [19], and RAPTOR [18] are all common WMRAs. Additionally, companion robots, robots on independent mobile platforms, have also been introduced. For example, UMASS/Smith ASSIST [19], Care-O-bot [20], El-E [21], COERSU [22], KARES II [23], WALKY [24], MOVAID [25], [26], HERMES [27]. For most of these tasks a WMRA, provides the best combination of cost effectiveness, convenience, and functionality for users. Thus, we focus on this robotic device for the remainder of the review.

## Wheelchair Mounted Robotic Arms

A typical wheelchair mounted robotic arm (i.e., WMRA or often called a robotic manipulator) is composed of multiple segments that comprise an "arm" attached to a base affixed to a user's wheelchair. The arm segments connect to an end-effector, gripper, or "hand" segment. This final segment may be used to grab and manipulate objects at a distance. These multiple segments require many degrees-of-freedom (e.g., 6 degrees of freedom may indicate 3 x, y, & z directions for translation and 3 yaw, pitch, roll directions of rotations for orientation) in order to move in a three-dimensional space. Common commercially available WMRA systems include the JACO robotic arm system developed by Kinova and the MANUS Assistive Robotic Manipulator (ARM) system created by Exact Dynamics [28].

Operation of such a WMRA may not be intuitive without extended practice or may be difficult for users who may have additional sensory and upper-body physical limitations from a disability. This has led to designing multiple control mechanisms for users to use which may include joysticks, touch screens, a single switch using hierarchical menus, eye-trackers, laser pointers, motion controllers, and others [28]-[29]. Further, many WMRAs may include detailed menus of complex hierarchies which may frustrate individuals with cognitive impairments in additional to a physical disability. Specifically, cognitive impairments such as decreased working memory or processing speed may impair individuals' ability to think through each hierarchical menu path. This has led to manufactures increasing the level of automation for assistive robots [30]. WMRAs have been designed to preform similarly to how a human would reach for and pick up an object. The trajectory of a human reach consists of two separate events: a large motion to get close to the object, followed by smaller movements to complete fine adjustments of the arm and hand [31]. These movements may be limited by the number of degrees of freedom of the robot, indicating that these two steps may depend on the specific robotic system. Broadly, an object retrieval task can be thought to consist of three parts: reaching for, grasping, and returning the object [29] – [32]. Additionally, cameras are often mounted on the WMRA, because users cannot always be expected to be able to use direct line of sight vision to see the objects they wish to retrieve or manipulate (e.g., on top of a bookshelf or under a table). Video or pictures of the robotic gripper or arm can be captured and displayed on a screen for the user. The automation of IDALs may also extend to the user's interface in an effort to provide easy access regardless of the disability. For example, using one click buttons [33] that allow for multiple steps or large movements to be completed at the same time. Additionally, this could extent to a single click, or press on a touch screen to a location. [29], or the user of a laser pointer to point to location in the real world [34]. Sensors have also been used to track users' speech, arm, hand, finger movements to generate gross motion up to an object [36, 37].

## Characteristics of Users of Assistive Robotic Technology

A large number of individuals with disabilities require assistance for moving around and interacting with their surroundings. The latest publicly available numbers from 2013 suggest that over 39 million community-resident Americans use assistive devices as mobility aids of which about a fourth use wheelchairs, scooters, or other mobility surrogates [38]. This is up from 6.8 million in the early 2000s [39]. Nearly one-third of mobility device users reported requiring assistance from other people in one or more IADLs, thus requiring, at least, a part-time caregiver. This is also a large concern outside the United States as well. A recent World Health Organization report estimated 15% of the world's population experience some form of disability [40]. Individuals with conditions that lead to mobility and upper-extremity limitations are those who may benefit the most from assistive robotic devices including WMRAs. Specifically, these conditions include neuromuscular deficits such as Spinal Cord Injury (SCI), Multiple Sclerosis (MS), Cerebral Palsy (CP), stroke, etc. Many of these individuals are confined to wheelchairs, have moderate to minimal function in their upper extremities, and require some amount of attendant care.

**Older adults**

Additionally, the elderly may also benefit from a robotic surrogate [41]. Prior research has shown older adults desire robotic assistance with maintaining their functional independence [42]. However, in the absence of specific disabilities, the IDALs, that older adults desire help with, may differ from individuals with disabilities. For example, older adults preferred robots to complete domestic maid type tasks such as cleaning and lawn care or to be social companion, but preferred to have a human assist with traditional IADLs such as meal preparation and cooking [7, 42, 43]. These differences may affect later user acceptance and use of a traditional assistive robot such as a WMRA. Additionally, age may confound traditional disabilities making the need for an assistive robot more salient. For example, 31% of people aged 80-84 require help with IADLs while this number jumps to 50% for people 85 and above [44]. This is a large and growing group with the elderly population. This becomes even more important as the number of older adults aged 65 and older is expected to double from 47.8 million in 2015 to 98.2 million by the year 2060 [45].

The monetary costs associated with attendant care are significant. For example, a 1990 national survey conducted by the Paralyzed Veterans of America estimated the average annual cost of a paid caregiver for SCI patients at $6,180 [46]. As an example of the cost of unpaid care, a recent survey of MS patients treated with disease modifying drugs in the United States estimated the average cost of care by family and friends at $4,614 annually with similar numbers across Europe [47, 48]. Besides these costs, there are other factors such as reduced privacy as well as the physical and emotional toll on the informal caregiver. Due to the large monetary costs or formal and the emotional and time cost of informal caregivers there is a substantial need for the development and rapid adoption of assistive robotic technology. Further, the use of robotic devices allows the user to regain control over their daily routine, increasing their functional independence. This is important to support their well-being. For example, one motivational theory called self-determination theory [49] stated that having autonomy over life choices and being able to complete tasks and goals is important for a positive sense of well-being. Thus, by utilizing robotic or quasi-robotic solutions, users can provide themselves with assistance to perform everyday activities while reducing caregiver dependence at the same time. This also allows for the socially reintegration of individuals with serious extremity limitations [39]. This is similar to how the modern powered wheelchair has alleviated mobility problems for individuals with moderately or severely limited lower extremity movement.

## Instrumental Activities of Daily Living

Users with many different conditions may need assistance completing ADLs due to impairments of their mobility [50]; however, these needs will vary greatly based on individuals' specific physical and sensory functional capabilities [51]. For example, the needs of an individual with a spinal cord injury will vary greatly based on the location of injury, and will also vary from the needs of an individual with ALS [8], Parkinson's disease, a healthy adult with a leg injury, or even an otherwise healthy frail older adult [51]. Even individuals with the same disability may vary in their functional capabilities [52] as no one disability presents exactly the same. Individuals' functional capability may also change over time due to factors such as treatment and further declines.

Broadly speaking however, individuals are likely to require assistance grooming, brushing one's teeth, bathing, preparing or eating food, watching TV [53], transferring in and out of a wheelchair, dressing and undressing [54], shopping, cleaning [55], completing errands, housework, meal preparation, telephone use, managing medications [7], or picking up dropped items. Overall, the tasks that can be completed using a WMRA can be grouped into three different categories (1) find and fetch, (2) pick and place, and (3) object manipulation. Examples of find and fetch tasks are tasks that involve finding an object in the environment, retrieving it, and then bringing it to the user such as brining a phone, drink etc. to the user so they can interact with it. Examples of pick and place tasks involve picking up an object from the environment and placing it at another location in the environment such as picking up a book from the floor to place it on the shelf, or picking up an item on the grocery store shelf and placing it into a shopping cart. On the other hand, object manipulation tasks involve interacting with an object in space which may include unlocking a door, assisting the user to put on make-up, or holding a phone to the user's ear and other tasks.

One might then ask which of these IADLs potential users would desire help completing on a day to day basis. Several studies have investigated this question. Several studies have shown potential users videos of an assistive robot completing different IDAL tasks then asked users to rate the utility and acceptance of the robot to provide assistance across a wide variety of IADL tasks. Of a wide variety of tasks surveyed, individuals with disabilities rated having robotic assistance completing IADLs involving a combination of find and fetch, pick and place, and object manipulation tasks would be highly useful (e.g., housework, meal preparation, completing errands, feeding [7, 56-57], etc. Results of another study focusing specifically on WMRA devices [58] surveyed a wide variety of users across different ages and disabilities who rated both find and fetch and pick and place tasks as essential IADLs they would require help with. More specifically, picking up objects from the floor and placing them on a shelf, table, or other location. Further, individuals also rated carrying objects were another critical IADL, robots with which a robot could assist. One longitudinal study, allowed individuals with spinal cord injuries in collaboration with their occupational therapists to use a WMRA for a several weeks. This study found users felt using a WMRA to assist with both find and fetch and pick and place IADLs would greatly increase their feelings of autonomy, competence, and quality of life [59]. Finally, [8] conducted an initial needs assessment and task analysis to determine which IADLs would have the most utility for elderly and disabled users. They followed a sample of individuals with IADLS for a week and recorded which ADLs users requested assistance for. Their results concurred with previous findings that object manipulation and retrieval tasks were among the most documented as being important to individuals with disabilities.

### User Acceptance of Assistive Robots

Robotic manipulators, can be complex, having multiple degrees-of-freedom. This complexity may indicate it may be more difficult for users to accept these devices. individuals with previous experience operating a professional service robot have been the ones most likely to indulge in interaction with a robotic assistant. Assistive robotic devices such as WMRAs present a unique challenge and an opportune testbed to study user acceptance of robots. Namely, users of assistive technology are a small segment of the overall population with limited access. Therefore, it is not surprising that there is little data to model human robot interaction for assistive robotics and the factors that affect user interaction.

Interaction with a robotic device that has multiple spatial degrees of freedom is complicated in itself even for the healthy population. It is even more complicated for disabled individuals where haptic feedback in the normal sense may not be possible due to sensory disabilities and the fact that the coordinates of visual feedback of the scene may not be correctly transformed into the coordinates of motor feedback due to visual perceptual or other cognitive disabilities. Furthermore, motor disabilities may prevent appropriate transfer of user intent in the absence of an appropriate user interface. As previously stated, many solutions have consequentially emerged in single-click interfaces and computer vision based controllers ([35, 36, 37]) that aim to spur performance by increasing robot autonomy. However, to the best of our knowledge, there do not exist many studies to indicate if user acceptance of assistive robots is on the rise commensurate with increases in performance and robot autonomy. Many examples of user research studies with assistive robotics have chosen to focus on performance and ignore the question of user acceptance. For example, one study developed different user interaction methods to use a WMRA in assisting users with several IDALs such as drinking with a straw, taking pictures of the environment and taking pictures of themselves (i.e., selfies), but focuses on an analysis of the robots' performance instead of the user's experience [28].

    To answer this question, a study was conducted with 10 spinal cord patients by the UCF Assistive Robotics Laboratory along with a local rehabilitation hospital [32]. All participants in this study were at least 21 years old, and 90 days post traumatic injury.

All patients were clinically diagnosed with an impairment due to damage to the spinal cord between C3 and C7. This means all of the patients demonstrated upper-body impairments depending on the location of their injury (i.e., neck, shoulders, arms, fingers; see [61] for a review). All participants also were required to have a power wheelchair and show sufficient cognitive ability to complete IADLs which was deemed equivalent to a Mini Mental Status Exam (MMSE – a test of cognitive impairment [62]) score of 23 or greater. Half of these individuals completed IADL tasks with a WMRA in fully automatic mode (Auto mode) where the users would select a target object and supervise the robot's performance, while the other half utilized a fully manual mode in which the participants would use a touch screen, trackball, microphone or jelly switch (depending on level of impairment) to control the robot. The latter condition required participants to move the arm and gripper (hand) segments separately using the interface requiring much more effort (Cartesian mode). Thus, Cartesian mode is the control mode in which users manually command 3-D Cartesian translation (forward, backward, left, right, up, and down) and 3-D rotation (yaw left, yaw right, pitch up, pitch down, roll clockwise, and roll counterclockwise) commands for the gripper at the end of the arm to enable interaction with the environment. On the other hand, Auto mode is the mode in which users click on an object of interest displayed on a screen as part of the scene captured by a gripper-mounted wide-angle video camera after which, the vision and control modules in UCF-MANUS automatically determine an end-effector trajectory to appropriately grab the object; this trajectory is executed when the user latches a switch. Thus, Auto mode users are not required to command multiple translation and rotation velocities at the end-effector.

     All participants used UCF-MANUS which is a modified version of the ARM robot from the University of Central Florida Assistive Robotics Laboratory. Mainly, visual information from a calibrated wide-angle stereo camera system mounted inside the gripper was processed by the robot while the visual information from one of the cameras was shown to the users. Additionally, pressure and touch sensors assisted the robot in avoiding collisions with environment and safely grasp objects while users used a multimodal user interface. Please see [36, 37, 63, 64]) for more details about the UCF-MANUS.

Both study groups across both conditions (Auto, Cartesian mode) completed were required to pick up 6 different objects (e.g., cereal box, remote, vitamin bottle) on one of two shelves (table height or ground level) to simulate common IADL tasks. The choice of objects and placement was designed to create a wide range of difficulty. Performance on each of these tasks was determined by several metrics. These include time on task (ToT) and the number and types of moves participants generated in each condition. These latter measures were used to generate a measure of planning and command inefficiency by comparing the participants performance to a model of optimal performance. Users' satisfaction was measured using the Psychosocial Impact of Assistive Devices Scale (PIADS [65]). The PIADS measure is designed to assess the impact of assistive technology on functional independence, quality of life, and well-being. It contains three subscales: Competence, Adaptability, and Self-esteem. Scores on each subscale range from – 3.0 (negative impact) to + 3.0 (positive impact).

The entire study was conducted over four weeks. The first week was comprised of pre-study assessments. In the second and third week participants completed a test without the robot, along with initial training on the UCF-MANUS, followed by specific training on all conditions. This training was used to ensure that all participants were sufficiently trained before their performance was analyzed. In the final week, participants used the UCF-MANUS to complete all tasks, followed by the PIADS.

The results of the study showed that, regardless of control mode, using the robot significantly increased individuals with spinal cord injuries' ability to complete IADLs (Figure 1). None of the patients could reach any of the objects on the bottom shelf without the robot and only approximately half of them could reach items on the top shelf placed to the left or right. However, with the robot, all individuals were able to complete all tasks representing a significant increase in their functional capability. Across control mode, the performance of the two cohorts did not significantly differ in the average amount of time it took to complete IADLs but the Cartesian mode required more effort as evidenced by number of interactions with the interface. Upon delving deeper into the results of the PIADS, an interesting observation was made. Although, using the MANUS robot in Auto mode increased subjects' functional capability and they had to use less effort than in the Cartesian mode, they were less satisfied with their experience than subjects participating in the Cartesian mode. In other words, the patients wanted to engage in manual control of the robot over the auto mode although it required more effort (Figure 2).

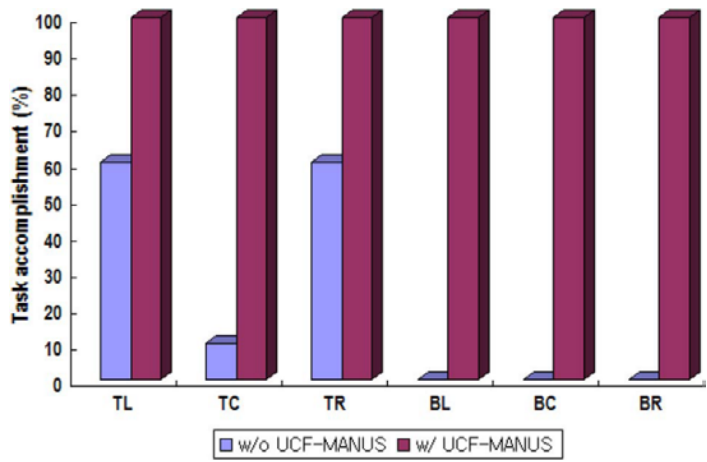

Figure 1. *Rate of task accomplishment with and without the assistive robot. [32]*

Although, it is difficult to generalize to everyone using only a small sample of users, these findings directly raise questions about the overall trend in robot automation. While a healthy able-body user may desire to off-load nuisance tasks to a personal robotic assistant or use them to meet social needs, this might not be the case for users with disabilities. This may represent a divergence between our research and that conducted with healthy older adults (e.g., [43, 49]). On the other hand, both may be explained through self-determination theory [7]. Everyone has needs for both having agency in their interactions with the world (autonomy needs) and being able to complete challenging goals (competence). Healthy users without any limitations have many outlets for how they meet these needs. However, as limitations start to impede individuals' ability to interact masterfully with the world, these outlets diminish. Thus, a desire for certain levels of autonomy may be directly proportional to an individuals'

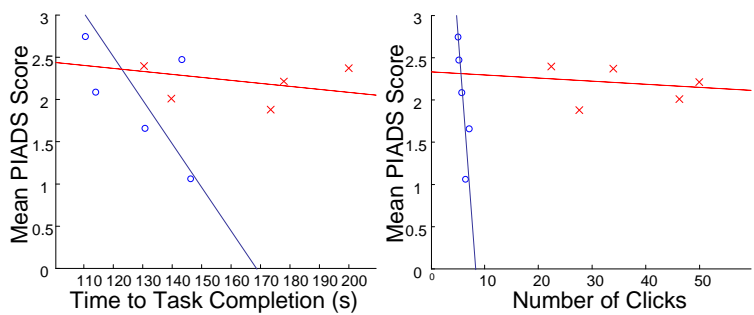

Figure 2. *User satisfaction is insensitive to performance when controlling robot manually (red x) while being highly sensitive to errors made by robot operating in autonomous mode (blue o) [32]*

ability to meet autonomy and competence needs. This means that older adults may show a greater desire to fully automate the tasks with which they have difficulty or dislike so that they can focus on those that are more likely to fulfil their needs. Because users with a disability are limited, manual control provides them with an outlet for regaining autonomy and competence.

**Individual differences In User Acceptance of Assistive Robotics**

Prior to the study in [32], we completed a focus group [108] with a potential group of participants. Each focus group member was a patient with spinal cord injury. This focus group found that potential users of WMRAs raised several issues of acceptance with this technology including: safety, simplicity, accessibility, customizability, reliability, and accuracy of the robot [59, 66] (Figure 3). In other words, users wanted a robotic system that: (1) would be effective and not cause injury or error (reliability, accuracy & safety), (2) was easy to use (simplicity), and (3) would be capable of adapting to their individual needs (accessibility and customizability). These issues map onto other models of technology acceptance. For example, reviews of the unified theory of acceptance and use of technology literature suggest factors that modify users' expectation of their performance or effort among others will impact how well users accept and in turn use novel technology [67, 68]. For users with disabilities, this may include factors such as their specific physical, sensory, or cognitive limitations associated with their disability and how well the assistive robotic device is designed to adapt to their specific needs. Additionally, how much choice users have over the robot's actions and how much control users have over how the robot completes its actions (i.e., self-determination theory needs) may also influence user acceptance. As already stated above, the dual-cohort pilot study in [32] found that while complete automation was the most efficient method for an assistive robot to complete an IADL, users preferred manually controlling the robot instead.

Further, these concerns map onto issues of robotic trust and usability of the human-robotic control interfaces. Robotic trust is the attitude (mindset/belief) that the robotic arm will help users complete their goals [69]. This attitude directly influences the willingness of users to both rely on information presented by the robot and to use or disuse the robot to implement that information. This attitude is shaped by users' attitudes, personality and prior experience. Individuals' level of trust will affect users' future behaviors as well as perceptions of both the robot and the task [70]. It is important to measure trust in the context of human-robot interactions to determine if users' trust with the system is properly calibrated as both over and under trust of a system may lead to error [71]. If trust is too low, users will disuse the system even when it could benefit them, and if trust is too high, users will over rely on the system and possibly exceed the robots limitations [69][71]. This it is important that the system takes this limitation of the user into account to prevent safety or performance issues.

Even when an assistive robotic device is completely controlled by the user, trust is a factor. This is because factors such as reliability, accuracy, and system capabilities are a part of trust of a complex technological system (e.g., [72, 73]). Thus, the term trust may be used in connection with user acceptance even in manual systems. However, trust becomes more of a concern when the robot and user act as a team working to complete a task (each with their own autonomy). In these cases, users would not have complete knowledge and must rely on the robotic system to some extent (e.g., [69, 74]). Thus, a usable control system for such a system would be designed with these user concerns in mind [66]. Further, these concepts have been applied to the acceptance of assistive robots. These studies show sociodemographic variables such as age, gender, previous experience with technology, education may also influence acceptance of assistive robotic devices and argue for an individualized approach [75]. These results align with previous surveys of potential users of assistive robotic devices [32].

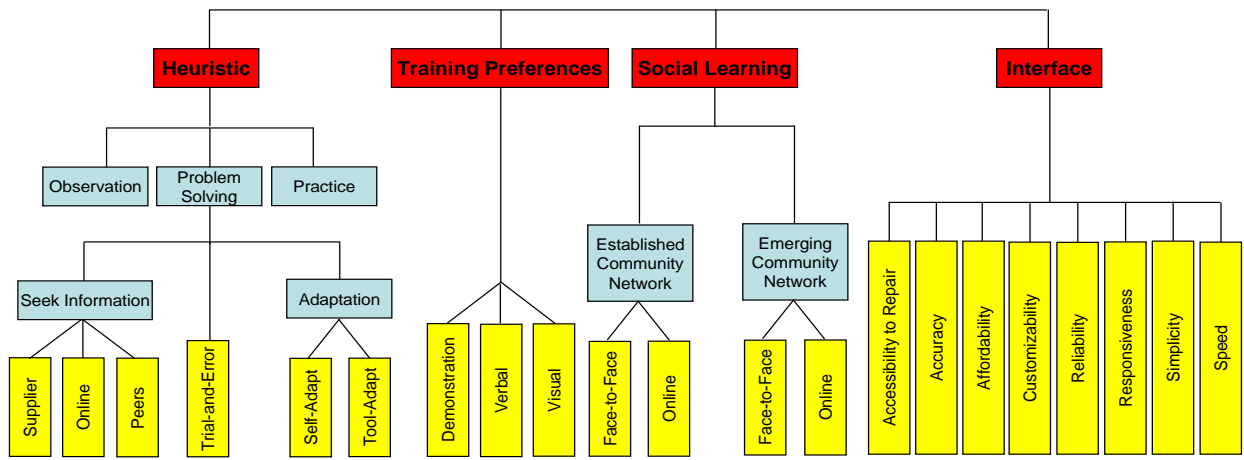

Figure 3. *Coding tree built through qualitative analysis of focus group data [107].*

## The Need for an Adaptive Human Robot Interaction (HRI) Framework for Assistive Robotics

Given the aforementioned gap in research for HRI for assistive robotic devices as well as the revealing results from previous research, there is a need for a framework for human robot interaction relevant to users with disabilities. Since users with disabilities report a more satisfying experience (as measured through psychometrics and corroborated by qualitative feedback measures) manually controlling and significantly interacting with a robot even when an autonomous robot shows markedly superior quantitative performance, interfaces must be designed to facilitate flexible HRI. While healthy users may prefer to cede autonomy to robots, we believe that disabled users tend to see the robot not merely as an agent to retrieve objects but also as a quintessential tool to reassert their domain of interaction with their environment as well as engage and exercise their cognitive faculties to the fullest. An autonomously acting robot is seen by this class of users as one more external agent that does their work for them. Thus, lesser reliance on a human caregiver does not result in a commensurate higher level of satisfaction when the interaction with the robot is sporadic and supervisory in nature. Based on this, there is a need for developing an adaptive HRI framework for users with disabilities where the autonomous capabilities of a robot may be harnessed as an enabling tool but not as the dominant overall control center.

It is well known that there is more variability in performance within the disabled population during HRI than there is within the healthy population. This implies that both the type of the interface and the level of interaction need to be readily customizable and scalable. While it is not entirely surprising that there would exist a variability across type of disability (e.g., MS versus SCI) and across diagnoses within a particular disability (e.g., SCI at the C3 vertebra versus at the C7 vertebra), analysis of the field trial data from [32] and the ensuing discussions with the occupational therapists (OTs) revealed a perhaps not so obvious user performance disparity even within a diagnosis. Furthermore, user performance also diverges along the time-of-onset (of injury/disease) axis owing to variations in type and length of rehabilitation, exposure to drug therapy regimes, job profile, and length of exposure to assistive technology among other factors. Thus, design of HRI based on initial diagnosis may not be appropriate. Therefore, an optimal framework must assess need for HRI based on objective measurements of user's performance and should be able to adaptively adjust the interaction to increase user satisfaction. In other words, an assistive robot needs to be calibrated or tailored for the specific limitations experienced by the individual user.

## Development of a Human Performance Model with Assistive Robotic Devices

In order to develop a framework for flexible/adaptive HRI, the relevant human performance characteristics (human factors) must be modeled for IADL tasks using WMRA devices. Previous work on robotic teleoperation and other human-robot interactions has defined many sensory, physical, and cognitive factors which may impact performance using a robotic device under manual control. These factors include visual perception (visual acuity & depth perception), psychomotor abilities, spatial ability (spatial orientation & visualization), and information processing ability (processing speed, memory span, & visual memory).

The first factor, visual perception encompasses both visual acuity and depth perception. Visual acuity is considered the sharpness or clarity of an individuals' vision, whereas depth perception, is a users' ability to utilize binocular cues to locate objects in the environment [76]. Both of these factors impact performance with a WMRA because they assist individuals to identify objects that are able to be grasped, as well as, interact with the robot. Furthermore, binocular cues are also important to human reach [77][78]. For example, the creation of an optimal motor trajectory that will result in a stable object grip [78]. When these cues are removed individuals compensate by increasing their grip aperture (i.e., they hold their finger and thumb farther apart) [78].

The second factor, psychomotor abilities refer to those characteristics allowing for manual manipulation. Deficits in psychomotor abilities can increase errors using the interface. Next, the third factor, spatial ability, is comprised of two components: spatial orientation based on an observer's point-of-view, and secondly, spatial visualization, viewpoint agnostic mental manipulation in a three-dimensional space. This factor is an important factor of HRI because it allows the user to create cognitive maps the workspace in order to make spatial decisions [80, 81]. For example, this may assist with integrating information from multiple camera angles, and their direct line of sight viewpoint. Spatial ability also works with other factors such as working memory, depth perception. For example, when making decisions regarding which item to grasp while trying to control the robot using the interface.

The final factor is information processing ability. This factor is comprised of processing speed, memory span, and visual memory. Processing speed refers to how fast an individual can process sensory, perceptual stimuli or retrieve information from long-term memory [82]. This ability declines with age [83]. Working memory capacity or span deals with how many items an individual can hold in memory at one time. Humans typically hold five to nine chunks of information at a single time, however, individual differences such as age and disability can affect memory span. Visual memory is the ability to remember the configuration, location, and orientation of visual material, and typically tests the short-term memory of presented visual material in iconic memory [82]. This factor is assists with remembering item detail when controlling the robot across multiple views [84]. These human factors were developed based on general robotic and teleoperation studies, because not many studies have examined user performance factors for assistive robotic technology including WMRAs. Given the importance of manual or semi-manual control modes, our previous work [85] developed a predictive model of how these human factors related to performance at controlling an assistive robotic manipulator.

In this previous study, 89 healthy participants, of both genders (45 men), were recruited between the ages of 18 – 63. Able-bodied participants were used in this initial study to limit the influence of factors other than the human factors measured. This study used the same modified robot as the study in [32]. Additionally, to reduce the influence of other confounding variables, all participants used the same graphical user interface displayed on a 12 x 9 monitor, controlled by a computer mouse. Prior to the study participants completed assessments designed to measure each human factor After completing these assessments, all participants were trained how to use the assistive robot. After a user reported being sufficiently trained, they were assigned 6 different pick and place and find and fetch tasks (e.g., reach an object from the top of a bookshelf). Each task varied in difficulty by varying the location or the number of steps required to reach the object. Performance was measured in this study using three measures: (1) time on task (ToT), (2) overall number of moves (NoMs), and (3) movement rate (NoM per minute; NoM/min). The results were analyzed using a combination of a machine learning categorization algorithm (Random Forest) and calculating several regression models. Overall findings from this research indicated that processing speed (RT), spatial ability (spatial visualization; V and spatial orientation; SO), working memory (WM), dexterity (gross; GD and fine dexterity; FD), and visual ability were predictive of performance. Visual acuity includes visual perception (VP), depth perception (DP), and visual acuity (both weak; WV and strong vision; SV) impacted performance across the different performance metrics (Figure 4).

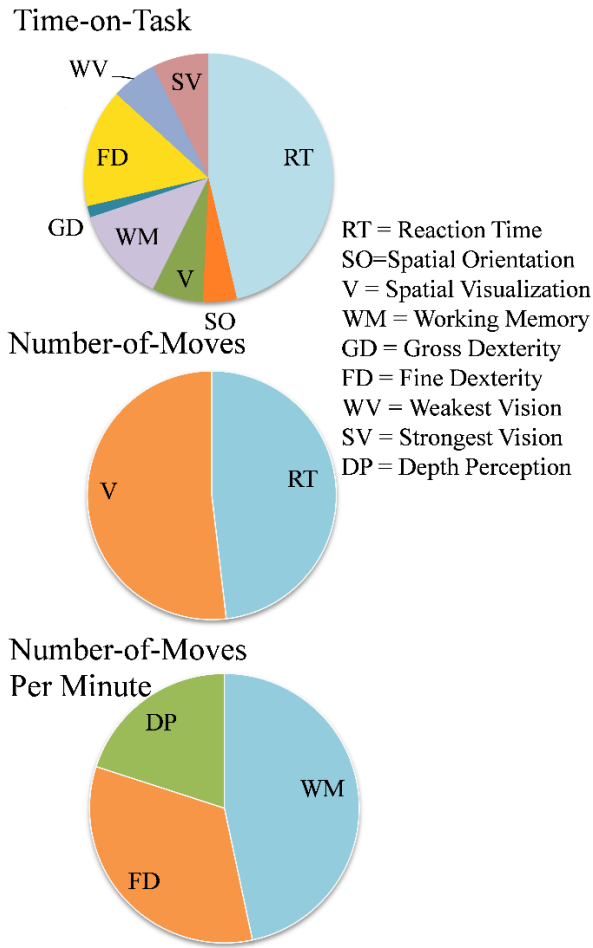

Figure 4. *Importance of different human factors for the three-performance metrics [85].*

The analyses in [85] suggested that different human factors were important to each of the different performance measures. RT, spatial abilities (SO & V), WM, dexterity (GD & FD), visual acuity (SV & WV) were all important factors for ToT; RT and V were important factors to the NoM; while WM, FD, and DP were all found to be important human factors to the NoM/min. These findings consolidate prior work that indicated spatial ability [80]-[81], WM [86], RT [86]-[87], and visual processing ability [25] were all important to both using a robot and performing manual reaching tasks.

Additionally, it was found that both GD and FD also predicted the completion of the experimental tasks. Dexterity is a critical biomechanical constraint to manual pointing and reaching tasks [88]-[91] and the findings of [85] extend this to include indirect HRI tasks which include reaching with a robotic manipulator. Findings for this factor may be contextually dependent on the interface used to interact with the robot and IADLs to be completed by the user. Thus, different interfaces may inherently (or via compensation) be more or less dependent on a user's dexterity.

**Age Differences in Performance Using a WMRA**

Because strength, dexterity, stamina, movement speed as well as fluid cognitive abilities (i.e., speed of cognitive processing, working memory efficiency, executive functioning, and attentional control) decline with age [92]-[95] when compared to younger adults. Because of these physical and cognitive differences, older adults may also interact differently with assistive robotic devices. To further investigate this question, the work in [94] utilized data from [93] to analyze age-differences in performance (Figure 5). The main findings from this work showed that older adults showed a longer ToT, and decreased movement rate than younger adults. Because the study found older adults performed significantly worse on measures of working memory capacity, spatial ability, and dexterity, these differences may have accounted for the performance differences found by the authors (Figure 6).

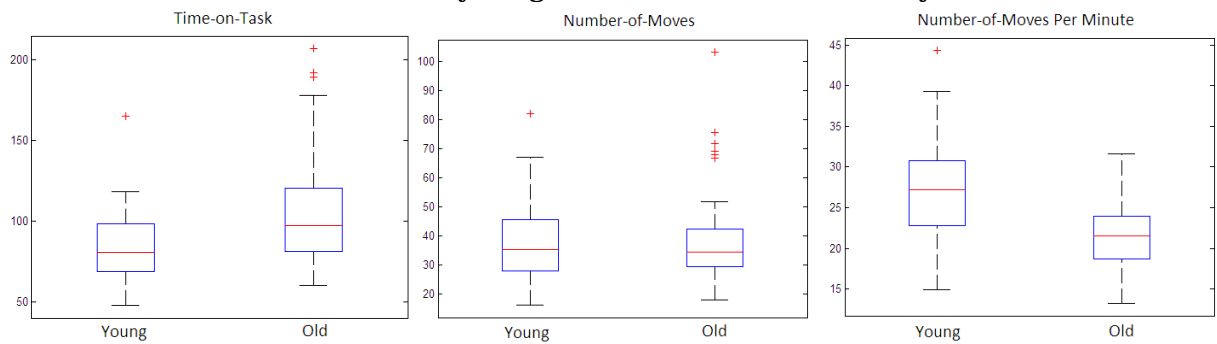

Figure 5. *Boxplots of performance metrics for younger and older cohorts showing the minimum, first quartile, median, third quantile, and maximum. Differences in ToT and NoM/min were seen to be statistically significant [96].*

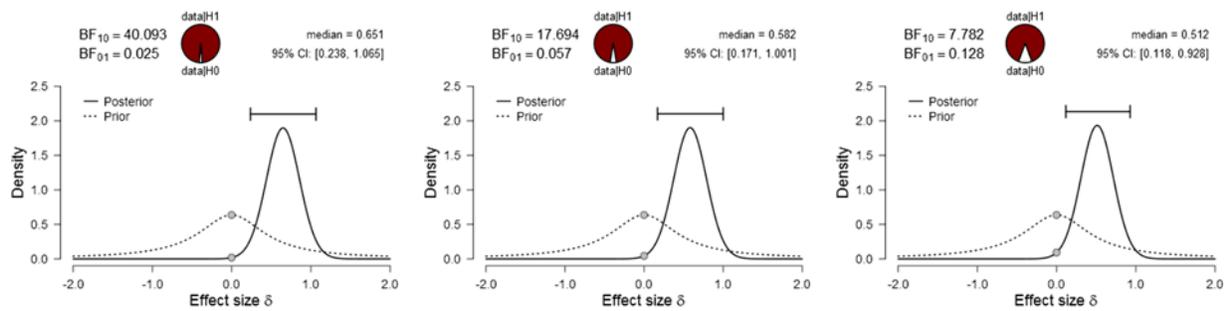

Figure 6. *Prior and posterior distribution of age-based differences in (a) Fine Dexterity, (b) Gross Dexterity, and (c) Working Memory [96].*

Additionally, the results may have also been found due to differences in movement and planning strategy between participant groups. However, the fact that performance differences were limited to an overall increase in ToT with no significant impact on NoM provides more evidence for an adoption of a cautious movement strategy than loss of cognitive processing abilities related to movement planning. In other words, the older cohort, experiencing declines in cognitive processing speed, proceeded with deliberation to arrive at a similarly efficient movement plan as the younger cohort. Further, regardless of ability differences found older adults still relied on information processing abilities (reaction time, working memory, and visual perception). This finding provided additional credence to the idea that older adults are implementing a more thought-out movement strategy. While younger adults are able to perform using their spatial abilities and some of their information processing ability (RT and WM), older users are using all facets of their information processing ability (VP in addition to RT and WM) as well as their spatial abilities [95].

**Gender Differences in Performance Using a WMRA**

Additionally, Paperno et al. (2018) [96] examined the effects of gender on performance with a WMRA as well. Unlike, performance differences due to differences in age gender differences are taught to be strategic and not due to a specific deficit. The only significant difference between men and women on human factors were found for dexterity. Specifically, women were found to have significantly better dexterity than men. In terms performance differences, women took longer to complete each task and had a greater number of moves. However, movement rate was similar between men and women (Figure 7 & Figure 8). These findings may suggest men and women navigated differently. Prior research has suggested men may have tried to pre-plan the entire route out before beginning the task while women may have broken the task into smaller segments, re-planning after each step [97]. Further, supporting this idea, Paperno et al. (2018) [96] found that while men relied more on WM, and spatial abilities (SO & V) women instead relied more on dexterity, while RT was important for both genders. Overall, these findings may suggest a need for an individualized approach for HRI based on age and gender differences between users. This prior research led to the selection of human factors relevant to the performance of WMRA. These factors in turn were also used to develop ways in which the robotic system may assist users through developing towards an adaptive HRI framework which includes how and when compensation should be developed with the needs of the user in mind.

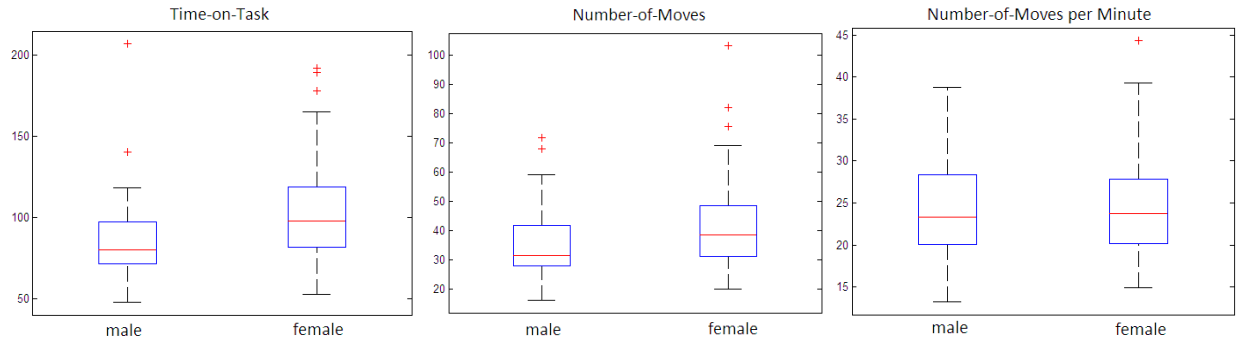

Figure 7. *Box plots of performance metrics for men and women showing the minimum, first quintile, median, third quantile, and maximum. Differences in TOT and NoM were seen to be statistically significant [96].*

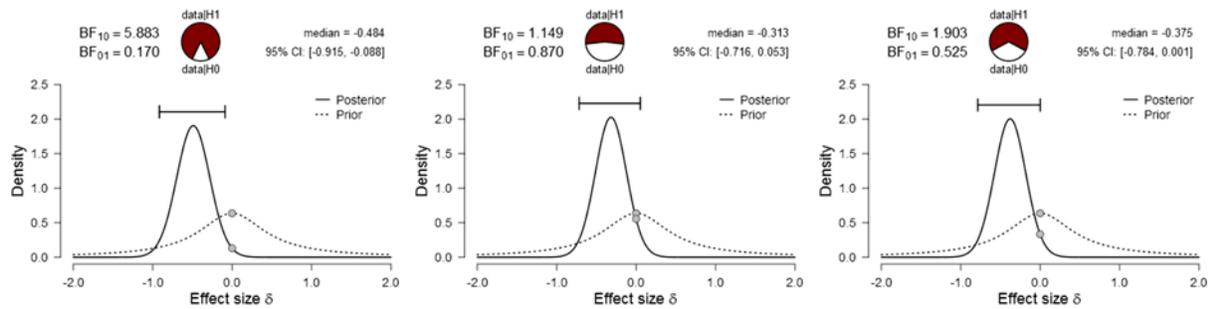

Figure 8. *Prior and Posterior Distribution of Gender-based Difference in (a) Gross Dexterity, (b) Reaction Time, and (c) Fine Dexterity [96].*

### Types of Human Robot Interaction, Levels, and Types of Automation

Parasuraman, Sheridan, and Wickens (2000) [98] developed an automation framework which described how automated devices such as robots may interact and take over tasks from human operators. This model defined four broad information processing stages: (1) sensory processing, (2) perception/working memory, (3) decision making, and (4) response selection and implementation. In short, the robot or automation may assist the user with information gathering, modify how it is presented to users, provide choices for action, and carry out tasks. Next, within each of these categories of processing stages lies various levels which describe the extent that the robot takes over each processing stage. These levels range from completely direct human-robot interaction (HRI) meaning the user manually controls the robot to completely automated and indirect HRI meaning the robot completely acts in place of the user. The robot's level of autonomy not only influences the division of labor of a specific task between the user and the automated system, but will also influence user acceptance of the robot along with the frequency that a particular task is automated. For example, is the specific type of automation employed by the system static, adaptable, or adaptive? In other words, does the robot always automate a task in the same way or does it respond to the user? Specifically, adaptive systems delegate the decision authority to the system while adaptable systems are set instead by the user [99]. Thus, in this framework, it is important for designers of systems to ask what, how, and how much should be automated with regard to a particular context of use [98,100,101].

This model is adaptable for users with disabilities because the limitations experienced by users may be explained in the context of Parasuraman et al.'s processing stages and implementation of an effective compensation strategy may be described in the context of varying the level of automation for a specific processing stage. Thus, while the operation of indirect robotic devices become more complex and increased automation is required a calibrated or adaptable approach rather than a simplistic off/on or adaptive method can be implemented to prevent a loss of user acceptance found in previous research [59, 66]. Further, this automation frameworks specifies many different ways to automate certain areas of the user experience while providing autonomy to the user.

While the question of how an assistive robot will compensate for the human factors developed by Paperno et al. (2016) [85] is still a subject of ongoing research it is possible to suggest some methods using current research into machine learning, computer vision, and psychological research [102]. It is also possible to consider adaptable ways in which the system can work along with the user to foster autonomy and competence. More obvious solutions exist for physical human factors such as dexterity and visual acuity. For example, because dexterity leads to issues performing gross and fine motor control, to compensate for poor dexterity, a range of control options could be used [103]). Interfaces using a graphical user interface (GUI) could manipulate button size so that less manual effort is required to select a target response. Further, the sensitivity of the input devices being used can be adjusted as well (at the response selection and implementation as the robot would change the way it responds to the environment). Further, physical devices such as a key guard could also be used to prevent the pressing of unintended keys on a touch screen [103].

Deficits in contrast sensitivity or visual acuity can be compensated by adding a multimodal feedback option, specifically one that uses sounds or larger text [104]. Additionally, a computer vision technique called object segmentation and detection may also be used to provide a user with a simplified display or magnify the field of view [102]. This technique could be used to implement color coding to maximize foreground to background contrast (see Jang, Lee, & Kim, 2016 [106] for an example of this technique). Further text recognition algorithms which can "read" text in an image (e.g., [107]) could also be used to highlight specific elements of a visual scene. Thus, these compensation strategies occur at the sensory or perception stage or Parasuraman et al. (2000)'s [98] framework because the input is modified before being presented to the user.

Deficits in working memory can be compensated for by reducing the amount of information that is needed to be stored at one time. Again, object segmentation may be useful to complete this task. Additionally, this can be accomplished through other changes to the interface, the use of movement shortcuts, or different levels of automation. Martens and Antonenko [108] stated that if there is adequate information provided through a GUI that is intuitive to use, it can lessen the effect of certain cognitive factors. Adaptation to the interface through extended experience or the use of intuitive design can also assist with decreasing working memory load [109]. As a user progressively uses the interface, the amount of automation for a specific task could be adjusted to maximize user performance while still allowing some level of manual control, thereby requiring less effort. This may also be used to help compensate for poor speed of processing ability. Finally, the user can also be provided with an overlay on a visual screen to suggest potential movement paths with ideal criteria (shortest ToT or NoM for instance). This compensation could be implemented after the use indicates the target location or object, then the system can provide guidance to the user. Previous research with users with disabilities such as cerebral palsy and multiple sclerosis experienced greater mental workload than healthy controls when using touch screen interfaces [109] which suggests a need to develop alternative interfaces for these users as well.

To implement this approach into a commercial system, users could work with the device manufacturer or a physical therapist to complete the testing of the relevant human factors. However, to create a more seamless approach, this testing may be able to be integrated directly into the system. For example, visual testing (visual acuity, contrast sensitivity, depth perception) could be completed online or through a computer application. Testing for spatial abilities, processing speed, working memory, and dexterity could also be automated as well. Applications for Android tablets are being developed to utilize short computer games to infer a user's score on certain human factors [110]. Another method may be through the use of inverse models. Paperno et al. (2016) [85] argued for the development of future technologies that could incorporate inverse models from the results of their research. Inverse models would not require specific testing but would infer a user's score on a specific human factor by the pattern of performance under manual control. These inverse models could then be used by the robot in question to predict which human factors the user has a deficit in and thereby implement appropriate compensators to reduce the effect of said deficits.

**Future Directions**

Future work should focus on design of hybrid interfaces that allow for easy transition among different control modes such as fully manual, fully autonomous, manual guided by robot/automation suggestions, and automated guided by human suggestions. For this to be implemented, sensors (vision, touch, hearing) need to be installed in the robotic system to understand both the scene and user intention; processing of this information can lead to a menu of choices in the human-robot interface that can be made available to the user to select. This menu of choices can be (a) generic based on a study of preferences of users interacting with robots, (b) tailored to a class of users based on known characteristics of such users, (c) tailored to characteristics of a known set of tasks/environments, or (d) adaptively tailored to a particular user and their environment based on a continuous study of interactions in the recent past. Finally, all design must give consideration to not only efficiency of human-robot interaction but also user satisfaction.